\documentclass[english,aps,twocolumn,showpacs,prl,color,superscriptaddress]{revtex4}
\usepackage[T1]{fontenc}
\usepackage[latin9]{inputenc}
\usepackage{graphicx}
\usepackage{babel}
\usepackage{color}
\begin{document}

\title{
Ca$_{2}$Y$_{2}$Cu$_{5}$O$_{10}$: the first frustrated quasi-1D ferromagnet close to criticality
}

\author{R.O.\ Kuzian}

\affiliation{Institute for Problems of Materials Science NASU, Krzhizhanovskogo
3, 03180 Kiev, Ukraine}
\affiliation{Leibniz-Institut f\"ur Festk\"orper- und Werkstofforschung IFW Dresden,
P.O. Box 270116, D-01171 Dresden, Germany}

\author{S.\ Nishimoto}

\affiliation{Leibniz-Institut f\"ur
Festk\"orper- und Werkstofforschung IFW Dresden,
P.O. Box 270116, D-01171 Dresden, Germany}

\author{S.-L.\ Drechsler} 
\thanks{Corr.\ author, E-mail: s.l.drechsler@ifw-dresden.de} 

\affiliation{Leibniz-Institut f\"ur
Festk\"orper- und Werkstofforschung IFW Dresden,
P.O. Box 270116, D-01171 Dresden, Germany}

\author{J.\ M\'alek}
\affiliation{Leibniz-Institut f\"ur
Festk\"orper- und Werkstofforschung IFW Dresden,
P.O. Box 270116, D-01171 Dresden, Germany}

\affiliation{Institute of Physics, ASCR, Prague, Czech Republic}
\author{S.\ Johnston}

\affiliation{Leibniz-Institut f\"ur
Festk\"orper- und Werkstofforschung IFW Dresden,
P.O. Box 270116, D-01171 Dresden, Germany}
\author{J. van den Brink}
\affiliation{Leibniz-Institut f\"ur
Festk\"orper- und Werkstofforschung IFW Dresden,
P.O. Box 270116, D-01171 Dresden, Germany}

\author{M.\ Schmitt}
\author{H.\ Rosner}

\affiliation{Max-Planck-Institute for Chemical Physics of Solids,
  01187 Dresden, Germany}

\author{M.\ Matsuda}

\affiliation{Quantum Condensed Matter Division, Oak
Ridge National Laboratory, Oak Ridge, TN 37831, USA}

\author{K.\ Oka} 

\affiliation{AIST, Tsukuba, Ibaraki 305-8562, Japan}

\author{H.\ Yamaguchi}

\affiliation{AIST, Tsukuba, Ibaraki 305-8562, Japan}

\author{T.\ Ito }

\affiliation{
National Institute of Advanced Industrial Science and Technology (AIST), Tsukuba, Ibaraki 305-8562, Japan}

\begin{abstract}
Ca$_{2}$Y$_{2}$Cu$_{5}$O$_{10}$ is build up from edge-shared CuO$_4$ plaquettes forming spin chains. 
From inelastic neutron scattering data 
we extract an in-chain nearest neighbor exchange $J_{1} \approx-170\,{\rm K}$ and the frustrating 
next neighbor $J_{2} \approx 32\,{\rm K}$ interactions, both significantly larger than previous estimates. 
The ratio $\alpha=|J_{2}/J_{1}|\approx0.19$ places the system very close to the 
critical point $\alpha_c=0.25$ of the $J_1$-$J_2$ chain, but in the {\em ferromagnetic} regime. 
We establish that the vicinity to criticality only marginally affects the
dispersion and coherence of the elementary spin-wave-like magnetic excitations, 
but instead results in a dramatic $T$-dependence of high-energy Zhang-Rice 
singlet excitation intensities. 
\end{abstract}


\date{\today}

\maketitle

Frustrated low-dimensional magnets serve as breeding grounds for novel and 
exotic quantum many-body effects.
Ca$_2$Y$_2$Cu$_5$O$_{10}$ (CYCO) and the closely related Li$_2$CuO$_2$ (LCO) 
are considered candidates for this type of unconventional and challenging physics
\cite{Matsuda01,Matsuda05}. 
These systems are in the family of frustrated edge-shared 
chain cuprates (ESC) and their magnetic excitation spectra, 
as probed by inelastic neutron scattering (INS), indicate striking puzzles.
It was claimed that the dispersion of the magnetic excitations in CYCO shows 
an anomalous {\it double} branch \cite{Matsuda01,Matsuda05}
while
LCO exhibits
a single, but weakly dispersing mode\cite{Boehm98}. 
These observations would point at a strong deviation of the dispersion 
from standard linear spin wave theory (LSWT), in any realistic 
ESC parameter regime. This motivates scenarios with
more sophisticated many-body physics, e.g.\ due to the presence of  
antiferromagnetic (AFM) interchain couplings (IC), causing the branch doubling 
in CYCO~\cite{Matsuda01,Matsuda05}. However, such a scenario invoking 
strong quantum effects seems to be at odds with the observed large and 
almost saturated magnetic moments $ \sim 0.9\mu_{\rm B}$ 
at $T\ll T_{\mbox{\tiny N}}=29.5$~K
\cite{Matsuda99,Fong99}, which suggests strongly {\it suppressed} 
quantum fluctuations. 
 To resolve the situation it is essential to identify the precise values 
of the exchange interactions in these ESC's, both within and 
between the spin chains. To this end it is key  
to measure and at the same time calculate the elementary 
magnetic excitations, ideally for directions of momentum transfer in 
which the excitations depend most sensitively
 on the strength of the in-chain couplings.
From scattering along the $a$-axis of CYCO, which does {\it not} fulfill 
the latter condition, a  moderate value of the ferromagnetic (FM)  
nearest neighbor (NN) coupling $J_1\approx-93$~K has been 
extracted \cite{Matsuda01} with a tiny frustrating  
AFM next-nearest neighbor (NNN) exchange $J_2\approx 4.7$~K (see Fig.\ 1).
\begin{figure}[b]
\includegraphics[clip,width=0.79\columnwidth]{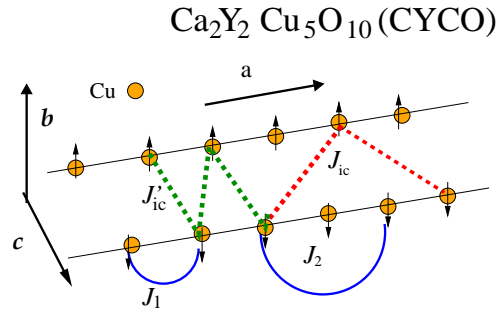}
\caption{(Color) Schematic view of the structure of the CuO$_2$ chain layer
and the main exchange paths of CYCO (see text).
}
\label{fig2} 
\end{figure}
From a theoretical point of view this is rather unexpected for the ESC 
chain geometry due to the presence of sizable O-O 2$p$ hopping along the chains. 
%
%
A recent reassessment of the 
exchange strengths based on INS on isotopically clean 
$^7$Li$_2$CuO$_2$ \cite{Lorenz09} has revealed a relatively large FM 
coupling $|J_1| > 200$~K, which is more than a factor 2 larger than earlier 
theoretical
values~\cite{Mizuno98}. Given the structural similarities to CYCO, one 
expects a larger value $J_1$ here too. In fact, 
high-$T$ $^{89}$Y-NMR data on CYCO appear difficult to reconcile with 
small $|J_1|$'s \cite{choi}.
Here we show that indeed by measuring with INS the magnetic excitations 
in CYCO along $\mathbf{Q}=(H,0,1.5)$, a direction where they 
are little affected by inter-chain couplings, one extracts a
 substantial  
in-chain $J_{1} \approx-170\,{\rm K}$ and 
a frustrating 
$J_{2} \approx 32\,{\rm K}$, so that $\alpha=|J_{2}/J_{1}|\approx0.19$. 
This indicates 
an exceptional position of CYCO within the ESC family: 
close to the critical point (CP) of the $J_1$-$J_2$ model $\alpha _c=1/4$, 
but on the FM side of its phase diagram and in contrast with   
Li$_2$ZrCuO$_4$ ($\alpha$= 0.3 \cite{Drechsler07}) and 
LCO ($\alpha=0.33$~\cite{Lorenz09}) 
which are on the {\it spiral} side of the critical point. We compare the obtained 
$J$'s  to a realistic 5-band extended Hubbard $pd$ model and 
L(S)DA+$U$ calculations, which are in good agreement. The resulting 
magnetic excitations calculated 
with exact diagonalization compare well to the ones 
obtained with LSWT, implying that the coherence of the elementary, 
spin-wave like, magnetic excitations are 
marginally affected by $\alpha$ and quantum fluctuations.
However, the relatively large $J_1$ 
and $\alpha$ values obtained
affect the thermodynamics \cite{Haertel11}.
The vicinity to the critical point
is probed by $\varepsilon=\alpha-\alpha_c$ 
strongly affects both the
magnitude and de- or increasing 
 $T$-dependence of the
Zhang-Rice singlet (ZRS) 
excitation intensity
for $\varepsilon > 0$ and $\varepsilon < 0$, respectively. 
This is manifest in  
resonant inelastic x-ray scattering (RIXS), EELS, and optics \cite{Malek08}
measurements as we will show.

CYCO  
has 
edge-shared CuO$_{2}$ chains 
along the 
$a$-axis.
The 
Cu$^{2+}$ spins 
are aligned
FM
along the 
$a$-axis (see Fig.\ 1). 
$ac$-planes
with CuO$_2$ chains alternate along the $b$-axis with
 magnetically inactive
cationic planes
containing
incommensurate 
and partially disordered 
CaY-chains 
which produce a non-ideal 
geometry in the 
CuO$_2$
chains. These mutual structural peculiarities
might be responsible for
the puzzling strong damping at large transferred neutron
momenta \cite{Matsuda01,Matsuda05} to be addressed elsewhere.

Our INS-study has  
been performed with a fixed final
neutron energy of 14.7~meV on a 3-axis neutron spectrometer TAS-2 installed
at the JRR-3
by the
Japan Atomic Energy Agency.
To analyze the 
dispersion 
of the 
magnetic excitations 
we adopt
the model given 
in Ref.\ \onlinecite{Matsuda01}. 
Then, 
CYCO
has the following main 
couplings $J(\mathbf{R})$, $\mathbf{R}\equiv(xa,yb,zc)$:
NN and NNN couplings along the chain $J(1,0,0)\equiv J_{1},$
$J(2,0,0)\equiv J_{2}$, and the interchain coupling (IC)
$J(0.5,0,0.5)\equiv J'_{ic},$ $J(1.5,0,0.5)\equiv J_{ic}$,
$J(0,1,0)\equiv J_{b}$=-0.06~meV, and 
$J(0.5,0.5,0)\equiv J_{ab}$=-0.03~meV. 
For the 
small interplane FM couplings $J_{b}$ and $J_{ab}$ 
we adopt the values 
from 
Ref.\ \onlinecite{Matsuda01}. 
Their contribution to the inchain dispersion
is negligible. 
Within the LSWT
the 
magnetic excitations 
dispersion 
 is given by Eq.\ (2) of Ref.\ \onlinecite{Matsuda01}:
$
\omega ^{2}({\mathbf{q}}) 
=
A_{\mathbf{q}}^{2}-B_{\mathbf{q}}^{2},\label{eq:wq} 
$
$
A_{\mathbf{q}} 
\equiv 
J_{\mathbf{q}}-J_{\mathbf{0}}+
\tilde{J}_{\mathbf{0}}-D, 
\quad 
B_{\mathbf{q}} 
\equiv 
\tilde{J}_{\mathbf{q}},\label{eq:gq}
$
where $J_{\mathbf{q}}=
(1/2)\sum_{\mathbf{r}}J_{\mathbf{r}}\exp\left(\imath\mathbf{q}\cdot\mathbf{r}\right)$ 
is the Fourier transform of intrasublattice interactions and analogously for 
the intersublattice interactions
$\tilde{J}_{\mathbf{q}}$.
The dispersion along 
$(0,0,L)$ 
shown in
 Fig.\ 3 (c) of 
Ref.\ \onlinecite{Matsuda01} depends
only on $J_{s}=J'_{ic}+J_{ic}$. Its value, as well as the 
anisotropy parameter $D$, may be found from the INS data for 
$\mathbf{q}=(0,0,0),(0,0,1.5)$:
$
J_{s}^{2}  =  \frac{1}{4}\left[\omega ^{2}(0,0,1.5)-\omega ^{2}(0,0,0)\right]
$, 
$
D = 2J_s-\omega (0,0,1.5).
$
%
Using $\omega (0,0,1.5)=5.03\pm0.03$, and $\omega (0,0,0)=1.63\pm0.01$~meV 
we obtain $J_{s}\approx$~2.35~meV, and $D\approx-0.27$~meV \cite{rem1}
which
is very
close to $J_{s}=2.24$~meV, and $D=-0.26$~meV found in 
Ref.\ \onlinecite{Matsuda01}.
\begin{figure}
\includegraphics[clip,width=0.82\columnwidth]{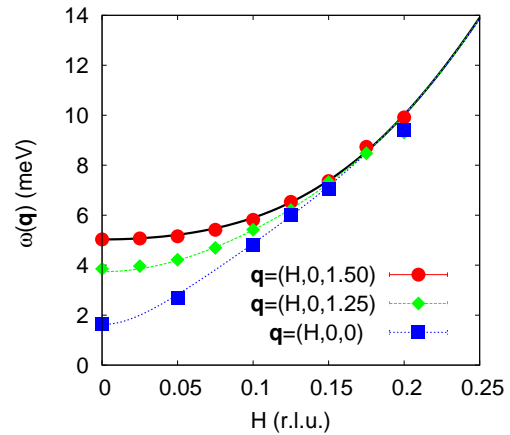}
\caption{(Color online) Dispersion along three
 lines of the first
 Brillouin zone parallel to the $a$-axis, 
 obtained from constant $\mathbf{q}$ scans.
The 
LSWT-fit was refined only for the dispersion along
(H,0,1.5) and it is 
shown by thin
lines.
}
\label{FIG2} 
\end{figure}
The dispersion along the line $(H,0,1.5)$
depends only on the
inchain $J$'s (if 
the tiny interplane  $J_{ab}=-0.03$~meV is 
ignored). It reads 
$
\omega (\mathbf{q})= A_{\mathbf{q}}=J_{\mathbf{q}}-J_{\mathbf{0}}+
\omega (0,0,1.5).
$
These $J$'s may be accessed from the dispersion along this line with 
much higher precision than from the previously reported data 
along $(H,0,0)$ 
and we have 
 \cite{rem1}:
\begin{eqnarray}
J_{1} & = & -14.69\pm0.5\ (4\%)\,{\rm meV}\approx-170.4\,{\rm K},\\
J_{2} & = & 2.78\pm0.2(7.6\%)
\,{\rm meV}\approx32.2\,{\rm K}.
\end{eqnarray}
The dispersions along 
$(H,0,0)$
(reported in 
Ref.\ \onlinecite{Matsuda01}) and $(H,0,1.25)$ (given here) are affected
by $J'_{ic}$ and $J_{ic}$. As we have determined only their sum, 
we adopt $J'_{ic}/J_{ic}=\tau \approx 4/9$ for the sake of concreteness
(suggested by our band structure
calculations). Notice that our $\tau$-value  
differs from 2 
adopted in Ref.\ \onlinecite{Matsuda01}. It 
might be refined empirically, if one measures also
along 
(H=1/6,K,L) for any K value. 

With the aim to detect quantum effects beyond the LSWT, 
we calculated  the
dynamical structure factor $S(\omega,q)$ 
using exact diagonalizations (see Fig.\ 3). 
In Fig.\ \ref{FIG2} the INS data together with the refined new
LSWT-fit are shown. 
$S(\omega,q)$ for our set
and that of Ref.\ 1 are
shown in Fig.\ 3.  The peak positions always nicely
follow the LSWT-curves, however our set
gives a better description 
of the INS-data than that in Ref.\ \onlinecite{Matsuda01}
(Fig.\ 4 therein) where the artificial double branching was ascribed to
AFM IC. Indeed, it
induces some intensity apart from the LSWT-curve, but 
these intensities are far too weak to be considered as 
a branch doubling. 
Notice the inflection point at $\pi/2$ for finite $\alpha $.
The total dispersion width is given solely by 2$| J_1| $.
\begin{figure}[b!]
\includegraphics[clip,width=0.76\columnwidth]{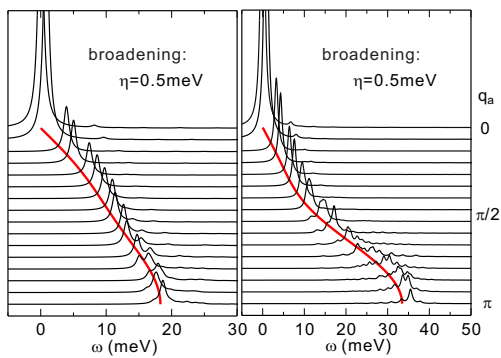}
\caption{(Color) Magnetic dynamical structure factor
$S(\omega,q)$ for the $J$-set 
of Ref.\ \onlinecite{Matsuda05} (left)
and for our
set 
(right) from exact diagonalizations
with $L=14\times 2$ and $15\times 2$ adopting 
$\tau=D=0$
and $J_{ic}=2.24$~meV, (see Ref.\ \onlinecite{Nishimoto11}.
Red line: dispersion from LSWT for the two parameter sets (see Fig.\ 2 and text).
}
\label{FIG3} 
\end{figure}

As an application of 
our data,
we consider
the 1D-magnetic susceptibility
$\chi(T)$ 
in the isotropic 
limit
(see Fig.\ 4).
\begin{figure}[t]
\includegraphics[width=0.6\columnwidth]{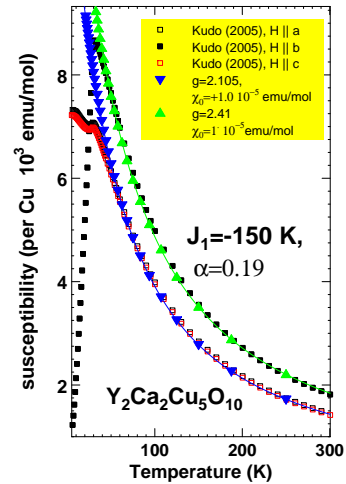}
\caption{(Color) Spin susceptibility $\chi(T)$ 
within
the isotropic
1D $J_1$-$J_2$ model using the 
transfer matrix renormalization group 
method, 
$\chi_0 $ - 
background susceptibility.
}
\label{figchi} 
\end{figure}
Despite the uncertainty caused by the unknown 
background susceptibility $\chi_0$,
the relatively large value
of
$| J_1 | $
doesn't allow to extract directly
$\Theta_{CW}$ from a 
$1/\chi(T)$ plot using only data up to 300~K. 
Instead a much broader $T$-interval up to about 
800~K would be required to reach the asymptotic high-$T$
limit
necessary for a proper quasi-linear behavior. Alternatively,
higher
orders
in the
high-$T$ expansion can be applied \cite{Schmidt11}.
Hence, 
the reported {\it AFM} values 
$\Theta_{CW}\sim -15$~K \cite{Yamaguchi99} 
or weak FM ones 
 $\Theta_{CW}\sim +8$~K \cite{Kudo05}
are rather artificial.
Using the $J$'s from our
INS-fits we predict instead a  markedly larger {\it FM}
value 
\begin{equation}
\Theta_{CW}=0.5(| J_1| -J_2) -J'_{ic}-J_{ic}-J_{ab}=+43.4\ 
\mbox{K}.
\end{equation}
Since we found $\alpha < \alpha_c$,
we readily
predict the value of the 2D
saturation field $H_s$ which is 
here
determined solely by the total AFM IC
like 
for
LCO
\cite{Nishimoto11}:
$
gH_s=4(J'_{ic}+J_{ic})=\frac{2}{g}77.4\ \mbox{T}\approx 64.8\ \mbox{T},
$
refining an estimate of 
70~T for $H \parallel b$ and $g=2.39$
from low-field magnetization data \cite{Kudo05}.
Next we consider the magnetic moment in the 
ordered state at low $T$.
Within the LSWT the reduction due-to quantum fluctuations
is about 6.8\% which yields 1.07$\mu_{\rm B}$ to be compared with the
experimental value of
0.92$\pm 0.08 \mu_{\rm B}$ \cite{Fong99} which however is affected
by the 
chemical reduction effect since about 0.22$\pm 0.04$
of the local moment resides on the O 2$p$ orbitals.

The exchange coupling strengths can also be determined by DFT+$U$ calculations.
%
For this we used the full potential scheme FPLO \cite{koepernik99} (vers.\ fplo9.01) and performed super cell calculations for different collinear 
spin arrangements applying LDA and GGA functionals \cite{pw,pbe}. 
The Coulomb repulsion $U_{3d}$ was varied in the physical relevant range 
from 5 to 8\,eV for a fixed  $J_{3d}=1$\,eV. 
In our calculations the 
incommensurate crystal structure of 
CYCO
can be treated only approximatively. 
Thus, we neglect (i) the modulation of the Cu-O distances within the 
CuO$_2$ chains 
and its buckling and (ii) the incommensurability of 
the CuO$_2$ and the
CaY subsystems. 
In particular,  the CuO$_2$ chains were treated as ideal planar chains 
reflecting an averaged
 Cu-O distance of 1.92~\AA\ and 
a Cu-O-Cu bond angle of 94.5$^\circ$. 
Furthermore, we modelled the CaY  layer by a Na
layer
 to preserve 
the half filling of the system.
The structure of the simplified model systems is given in
Ref.\ \onlinecite{supp}.
These structural
simplifications 
allow a reliable modelling
 of $J_1$ yielding an FM  value: $\sim$ -150~K~\cite{supp}.
In previous studies of closely related ESC \cite{li2zrcuo4,linarit}
for  the 
effects
of (i) chain buckling and (ii) the cation related crystal fields 
it was shown that $J_1$ is rather robust 
whereas $J_2$ is 
strongly
reduced 
by a factor 2 to 3. Thus,  our  $-J_1
\sim 150~K$ is 
considered as a rather reliable lower estimate by  about 10 to 20\% with 
respect to the 
buckled chain geometry in CYCO. 
However,  
in view of
the drastic dependence of $J_2$ on these 
parameters \cite{li2zrcuo4,linarit}, 
a derivation of a reliable value from the applied model structure is 
difficult.
Thus, CYCO fits the general experience 
of a sizable FM $J_1$-value for ESC in 
contrast to the assignments of only a few K, 
proposed 
for LiVCuO$_4$
\cite{Enderle}
and NaCu$_2$O$_2$ \cite{Capogna}.
Such  small $J_1$'s would put them in a region of strong quantum fluctuations, 
harboring the difficulty that 
the observed pitch angle cannot be described classically \cite{NishimotoEPL}.

Whereas the vicinity of CYCO to the quantum critical point only weakly 
affects the dispersion and coherence of the elementary, spin-wave like, 
magnetic excitations, we will show that the amplitude to excite  
Zhang-Rice singlets (ZRS) at 
typical high-energies, probed
by spectroscopic means depends strongly on the 
frustrating $J_2$.
and temperature.
We have also performed exact diagonalization 
calculations using an extended 5-band Cu 3$d$ O$2p$ 
Hubbard model for CYCO with a standard
parameters \cite{remarkparameters}.
To fit the INS-derived value of $J_2=$32~K, the $t_{px,px}=0.59$~eV
has been slightly reduced as compared with LCO (0.84~eV, $J_2=$~76 to 66~K) 
which simulates probably the deviations from the ideal chain geometry.
Mapping the spin states of the 5-band Hubbard model
onto a frustrated spin model,
we obtain 
$J_1=-177.5$~K and $J_2=32.3$~K in full 
accord
with the LSWT-analysis of the INS data given above.
We stress that the value of $J_1$ is mainly determined by the
direct FM
 exchange coupling $K_{pd}=65$~meV and not
by the Hund's rule coupling on the O sites $J_p=0.5\left( U_p-U_{p_xp_y}\right) 
=0.6$~eV
as often adopted. The 
significant value of $J_1$
is generic for ESC with a Cu-O-Cu bond angle 
$\varphi <
96^\circ$ at variance from the 
case of CuGeO$_3$
with $\varphi \approx  98^\circ$ causing an AFM $J_1$.
\begin{figure}[t]
\includegraphics[width=0.85\columnwidth]{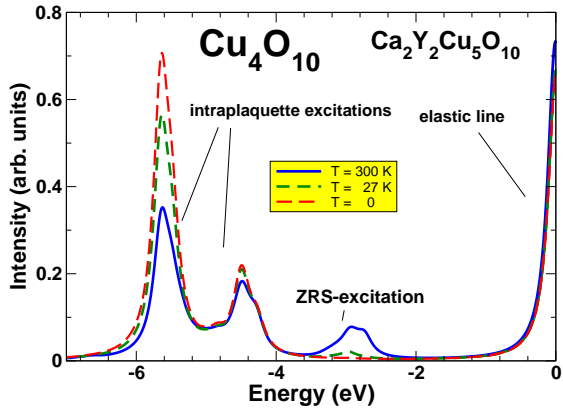}
\caption{(Color) 
$T$-dependent O-K RIXS-spectrum for  
$xx$-polarization 
from a Cu$_4$O$_{10}$ cluster within exact diagonalization
using 
a Lorentzian
broadening (half width $\Gamma=0.13$~eV).  
}
\label{figoptzrs} 
\end{figure}

As a modern spectroscopy, RIXS provides valuable insights  
into the correlated orbital and electronic structure
(for a review see Ref.\ \onlinecite{Ament2011}). Therefore, we also studied 
the $T$-dependent O-K edge RIXS-spectra for a Cu$_4$O$_{10}$ cluster within 
exact diagonalization (see Fig.\ 5).  We find a strong 
{\it decrease} of the intensity for the ZRS excitons
with decreasing $T$, which is qualitatively
in accord with general considerations for EELS and optics
\cite{Malek08}.
The not yet assigned
feature observed for CYCO 
at 300~K near 527~eV (Fig.\ 5 in Ref.\ \onlinecite{Kabasawa05},  
counted from the 530~eV excitation energy) corresponds just to these excitations. 
It perfectly agrees
with 2.95~eV obtained here. 
Notice
that when one sets $t_{px,px}=0$, thereby 
strongly suppressing $J_2\propto t^2_{px,px}$, no ZRS exciton is observed even
at 300~K. Hence, 
the frustrated FM 
differs 
qualitatively from a pure FM 
in its high-energy response.

To summarize, we have shown that Ca$_2$Y$_2$Cu$_5$O$_{10}$ is a 
frustrated quasi-1D ferromagnet close to criticality. This 
edge-sharing spin-chain compound has pronounced FM correlations in the 
presence of a sizable in-chain frustration. 
The main intensity of magnetic excitations in $S(\omega,q)$ is reasonably well 
described within LSWT. The signatures of the sizable in-chain 
frustration found here cause (i) a characteristic curvature of
the in-chain dispersion of magnetic excitations and (ii) Zhang-Rice singlet 
features at $\sim 3$~eV that are 
strongly $T$-dependent and are detectable by spectroscopies. 
%
The spin alignment between the chains below $T_N$ is supported 
by a specific AFM inter-chain coupling which determines the saturation 
field. It causes interesting quantum effects that go beyond a linear spin wave description. 
These effects, and the interplay of  $J'_{ic}$ and $J_{ic}$, provide an interesting 
problem not yet investigated in full detail. However, such quantum
effects are neither strong enough to cause a breakdown of LSWT, nor to 
induce an additional branch of magnetic excitations, as suggested previously. Our results 
can further aid in  
the correct assignment
of the frustration \cite{Enderle,Capogna} in 
 other ESC, including multiferroics, but especially for the complex chain-ladder system
 (La,Sr,Ca)$_{14}$Cu$_{24}$O$_{41}$ \cite{carter96} which also possibly contains 
 frustrated FM CuO$_2$ chains as suggested by the 
  edge-sharing geometry 
  and sizable NNN transfer integrals \cite{schwing07}
  giving rise to significant AFM
  $J_2$'s. 
  Its nonideal chains, as with CYCO,
  challenge one to look for more sophisticated but yet 
  solvable theoretical models that include
  incommensurate and disorder effects.

\vspace{-0.15cm}  

We thank the DFG for financial support. 
In particular, the Emmy-Noether program is acknowledged for founding.
 We also acknowledge fruitful discussions with J.\ Richter, J.\ Geck, 
 and V.\ Bisogni.

\begin{widetext}
\centerline{
\large \bf EPAPS supplementary online material:}
\centerline{\large \bf 
 \char`\"{}
Ca$_{2}$Y$_{2}$Cu$_{5}$O$_{10}$: the first 
  frustrated quasi-1D ferromagnet close to criticality
\char`\"{}
}

\vspace{0.3cm}
\noindent
R.O.\ Kuzian, S.~Nishimoto, S.-L.~Drechsler, J.\ M\'alek, J.\ van den Brink\\
IFW Dresden, P.O.~Box 270116, D-01171 Dresden, Germany\\
M.\ Schmitt, and H.\ Rosner\\
Max-Planck Institute for Chemical Physics of Solids, 01187 Dresden,  Dresden, Germany\\
M.\ Matsuda,\\
Quantum Condensed Matter Division, Oak Ridge National Laboratory, Oak Ridge, TN 37831,
USA\\
K.\ Oka, H.\ Yamaguchi and T.\ Ito\\
AIST, Tsukuba, Ibaraki 305-8562, Japan\\
\vspace{0.7cm}

\renewcommand{\theequation}{S\arabic{equation}}
\renewcommand{\thefigure}{S\arabic{figure}}
\renewcommand{\thetable}{S\arabic{table}}
\setcounter{equation}{0}
\setcounter{figure}{0}
\setcounter{table}{0}

{\small
In the present EPAPS 
supplementary online material we present 
(i)
exact diagonalization
studies of the dynamical structure factor for coupled chains with various interchain 
couplings of the type as realized in CYCO.
(ii) We show our prediction and details of the calculated optical
conductivity $\sigma(\omega)$
(iii) We demonstrate the influence of interchain coupling and spin anisotropy
on the magnetic susceptibility.
(iv) we provide the reader with further details of the LSDA+$U$ 
and GGA+$U$ 
calculations.
(v) we illustrate the influence of inchain frustration on the dispersion of
magnetic excitations focussing on its curvature behavior and in particular
on the positions of inflection points 
within linear spin-wave theory.}\\

\vspace{0.2cm} 
\subsection*{The spin-Hamiltonian with uniaxial
anisotropy and CYCO geometry}
In general, spin waves are the quanta of small oscillations of
spins around the classical ground state of the spin-Hamiltonian. So,
for the same Hamiltonian, the form of the dispersion may
be radically different for different values of parameters, since
the classical ground state may be different from the quantum one. 
The spin-Hamiltonian
for CYCO reads 
\begin{eqnarray}
\hat{H} & = & \hat{H}_{A}+\hat{H}_{B}+\hat{H}_{AB}\label{H}\\
\hat{H}_{A(B)} & = & \frac{1}{2}\sum_{\mathbf{m}\in A(B)}
\left\{ \sum_{\mathbf{r}}
\left[J_{\mathbf{r}}^{z}\hat{S}_{\mathbf{m}}^{z}\hat{S}_{\mathbf{m}+\mathbf{r}}^{z}
+J_{\mathbf{r}}\hat{S}_{\mathbf{m}}^{+}\hat{S}_{\mathbf{m}+\mathbf{r}}^{-}
\right]\right. \nonumber \\
 & + & \left.\sum_{\mathbf{R}}
 \left[J_{\mathbf{R}}^{z}\hat{S}_{\mathbf{m}}^{z}\hat{S}_{\mathbf{m}+\mathbf{R}}^{z}
 +J_{\mathbf{R}}\hat{S}_{\mathbf{m}}^{+}\hat{S}_{\mathbf{m}+\mathbf{R}}^{-}
 \right]\right\} \label{eq:HA} \\
\hat{H}_{AB} & = & \sum_{\mathbf{m}\in A}\sum_{\mathbf{f}}
\left[\tilde{J}_{\mathbf{f}}^{z}\hat{S}_{\mathbf{m}}^{z}
\hat{S}_{\mathbf{m}+\mathbf{f}}^{z}
\right. \nonumber \\
 & + & \left.
\frac{\tilde{J}_{\mathbf{f}}}{2}\left(\hat{S}_{\mathbf{m}}^{+}
\hat{S}_{\mathbf{m}+\mathbf{f}}^{-}+
\hat{S}_{\mathbf{m}}^{-}\hat{S}_{\mathbf{m}+\mathbf{f}}^{+}
\right)\right]\label{eq:HAB}
\end{eqnarray}
 where $\mathbf{m}$ enumerates the sites in one sublattice, ($\mathbf{r}=\pm n\mathbf{a},\: n=1,2\ldots$)
determines the neighboring sites within the chain, $\mathbf{a}$ being
the lattice vector along the chain, vector $\mathbf{f}$ connects
the sites of different chains in the $\mathbf{ac}$ plane, $\mathbf{R}$
connets the sites in different planes, but in the same sublattice
in CYCO. 
We have allowed for an uniaxial
anisotropy of the exchange interactions. 
\subsection{Linear spin-wave theory}
For the collinear AFM the classical ground state is the
Ne\'el state, the spins on the $A$ sublattice are directed up, and down
in the 
$B$ sublattice. We introduce two different sets of spin-deviation
operators
\begin{eqnarray}
\hat{S}_{\mathbf{m}\in A}^{+} & \equiv & \sqrt{2S}f_{\mathbf{m}}(S)a,
\ \hat{S}^{-}\equiv\sqrt{2S}a^{\dagger}f_{\mathbf{m}}(S),\label{SA}\\
\hat{S}^{z} & \equiv & S-\hat{n}_{\mathbf{m}},\ 
\hat{n}_{\mathbf{m}\in A}  =  a_{\mathbf{m}}^{\dagger}a_{\mathbf{m}}\label{nmA}\\
\hat{S}_{\mathbf{m}\in B}^{-} & \equiv & \sqrt{2S}f_{\mathbf{m}}(S)b,
\ \hat{S}^{+}\equiv\sqrt{2S}b^{\dagger}f_{\mathbf{m}}(S),\label{eq:SB}\\
\hat{S}^{z} &\equiv & -S+\hat{n}_{\mathbf{m}},\ 
\hat{n}_{\mathbf{m}\in B}  =  b_{\mathbf{m}}^{\dagger}b_{\mathbf{m}},\label{nmB}\\
f_{\mathbf{m}}(S) & = & \sqrt{1-\hat{n}_{\mathbf{m}}/2S} \\
&=& 1-\left(\hat{n}_{\mathbf{m}}/4S\right)-
\frac{1}{32}\left(\hat{n}_{\mathbf{m}}/S\right)^{2}+\cdots\label{eq:fm}
\end{eqnarray}
 The Ne\'el state (AFM ordering of FM chains
in the $\mathbf{ac}$ plane) is the vacuum state for the operators
$b\left|\mathrm{Neel}\right\rangle =0$. So, the operator $b(a)$
annihilates the spin deviation from the Ne\'el 
order (which means a
spin-flip for $s=1/2$) on the sublattice $B(A)$. The Hamiltonian
(\ref{H}) can be rewritten as 
\begin{eqnarray}
\hat{H} & = & \hat{H}_{0}+\hat{H}_{int},\label{Hviab}\\
\hat{H}_{0} & = & \sum_{\mathbf{q}}
\left[A_{\mathbf{q}}\left(a_{\mathbf{q}}^{\dagger}a_{\mathbf{q}} 
+ b_{\mathbf{q}}^{\dagger}b_{\mathbf{q}}\right) \right. \nonumber \\
&+& \left. B_{\mathbf{q}}\left(a_{\mathbf{q}}b_{-\mathbf{q}}
+a_{\mathbf{q}}^{\dagger}b_{-\mathbf{q}}^{\dagger}\right)\right],
\label{H0}\\
A_{\mathbf{q}} & \equiv & 
S\left[\sum_{\mathbf{r},\mathbf{R}}J_{\mathbf{r}}\exp\left(\imath\mathbf{qr}\right)-
\sum_{\mathbf{r},\mathbf{R}}J_{\mathbf{r}}^{z}+
\sum_{\mathbf{f}}\tilde{J}_{\mathbf{f}}^{z}\right],\label{eq:eq}\\
B_{\mathbf{q}} & \equiv & 
S\sum_{\mathbf{f}}I_{\mathbf{f}}\exp\left(\imath\mathbf{qf}\right)  \quad .
\label{eq:gq}\end{eqnarray}
 The Fourier transform of e.g.\ 
  $b_{\mathbf{q}}$ reads $b_{\mathbf{q}}=\sqrt{2/N}\sum_{\mathbf{m}\in B}\exp(-\imath\mathbf{qm})b_{\mathbf{m}}$;
$N$ denotes the total number of sites. The transverse part of $\hat{H}$
(\ref{H}) defines the one-particle hoppings in $\hat{H}_{0}$ (\ref{H0}),
the Ising part contributes on-site energy values. The terms, which
contain more than two spin-wave operators enter $\hat{H}_{int}$.
The magnons $\alpha_{\mathbf{q}},\beta_{\mathbf{q}}$ in an AFM
are introduced as a mean-field solution of Eq.\ (\ref{Hviab}) \cite{Oguchi60},
which \emph{neglects} 
$\hat{H}_{int}$. 
\begin{eqnarray}
\left[\alpha_{\mathbf{q}},\hat{H}\right] & \approx & 
\omega_{\mathbf{q}}\alpha_{\mathbf{q}},\;
\left[\beta_{\mathbf{q}},\hat{H}\right]\approx
\omega_{\mathbf{q}}\beta_{\mathbf{q}}\label{eq:mf}\\
\omega_{\mathbf{q}} & = & \sqrt{A_{\mathbf{q}}^{2}-
B_{\mathbf{q}}^{2}},\label{eq:wSW}\\
\alpha_{\mathbf{q}} & = & \cosh\theta_{\mathbf{q}}a_{\mathbf{q}}+
\sinh\theta_{\mathbf{q}}b_{-\mathbf{q}}^{\dagger},\label{eq:uv}\\
\beta_{\mathbf{q}} & = & \cosh\theta_{\mathbf{q}}b_{-\mathbf{q}}+
\sinh\theta_{\mathbf{q}}a_{\mathbf{q}}^{\dagger}\nonumber \\
\tanh2\theta_{\mathbf{q}} & = & B_{\mathbf{q}}/A_{\mathbf{q}}.\nonumber 
\end{eqnarray}
Now we specify the expressions for $A_{\mathbf{q}},B_{\mathbf{q}}$,
which result from the geometry of CYCO
and the spin value $s=1/2$. In the summations over $\mathbf{r},\mathbf{R},\mathbf{f}$
we retain only the following terms. For the in-chain exchanges we
retain $J_{1},\: J_{2}$, which correspond to $\mathbf{r}=\mathbf{a},\:2\mathbf{a},$
respectively. For the interchain interactions we retain $J_{ic}^{\prime},\: J_{ic},\: J_{b},\: J_{ab}$,
which correspond to $\mathbf{f}^{\prime}=\left(\mathbf{a}+\mathbf{c}\right)/2$,
$\mathbf{f}=\left(3\mathbf{a}+\mathbf{c}\right)/2$, 
$\mathbf{R}_{b}=\mathbf{b}$,
$\mathbf{R}_{ab}=\left(\mathbf{a}+\mathbf{b}\right)/2$.
Then we find
\begin{eqnarray}
A_{\mathbf{q}} & = & 
J_{1}\left(\cos q_{a}a-1\right)+J_{2,}\left(\cos2q_{a}a-1\right) \nonumber \\
&+& J_{b}\left(\cos q_{b}b-1\right)+\nonumber \\
 & + & 2J_{ab}\left(\cos\frac{q_{a}a}{2}\cos\frac{q_{b}b}{2}-1\right)+
 2
 J_s-D,\label{eq:eqli}\\
B_{\mathbf{q}} & = & 2\left[J_{ic}^{\prime}\cos\frac{q_{a}a}{2}
+J_{ic}\cos\frac{3q_{a}a}{2}\right]\cos\left(\frac{q_{c}c}{2}
\right).\label{eq:gqli}\end{eqnarray}
 Note that interplane interactions enter $A_{\mathbf{q}}$ as they
represent intra-sublattice interactions. The value of the gap at 
$\mathbf{q}=0$
is
\begin{equation}
\Delta  =  \sqrt{A_{0}^{2}-B_{0}^{2}},
\end{equation}
 the anisotropy enters the dispersion via the integral value $D$,
which is defined as 
\begin{equation}
A_{0}-B_{0}\equiv D=\sum_{\mathbf{f}}\left(\tilde{J}_{\mathbf{f}}^{z}-
\tilde{J}_{\mathbf{f}}\right)-\sum_{\mathbf{r},\mathbf{R}}
\left(J_{\mathbf{r}}^{z}-J_{\mathbf{r}}\right),\label{eq:Ddef}
\end{equation}
 and is related to the gap value by the relations
 \begin{eqnarray}
\Delta & = & \sqrt{D\left(D+2B_{0}\right)},\nonumber \\
D & = & \left(\sqrt{B_{0}^{2}+\Delta^{2}}-B_{0}\right).\label{eq:Dval}
\end{eqnarray}
\subsection*{Dispersion of magnetic excitations and inchain frustration}
Ignoring tiny interactions like $J_b$ for $qc=2(m+1)\pi$, $m= 0,1,2, ...$
the dispersion along (H,0,1.5)
within the LSWT reads (see Eq.\ 2 in Ref. \onlinecite{Matsuda01}):
\begin{eqnarray}
\frac{\omega}{\mid J_1 \mid}&=&   1-\cos x +\alpha \left( \cos 2x -1\right) + \nonumber \\ 
&& 2\beta +\gamma \left( \cos 3x -1 \right) + ...,  \ ., \nonumber \\
\end{eqnarray}
where $x=qa$, $\gamma= J_3/\mid J_1 \mid $ denotes the third neighbor
inchain coupling.  The dimensionless 
interchain coupling constant slightly "renormalized"
by the anisotropy parameter $D$: 
$\beta_D
=(J_s-0.5D)/\mid J_1 \mid \approx 0.17$ 
has been omitted
in Fig.\ S1 since it doesn't affect the curvature
\begin{figure}[b]
\includegraphics[width=0.75\columnwidth]{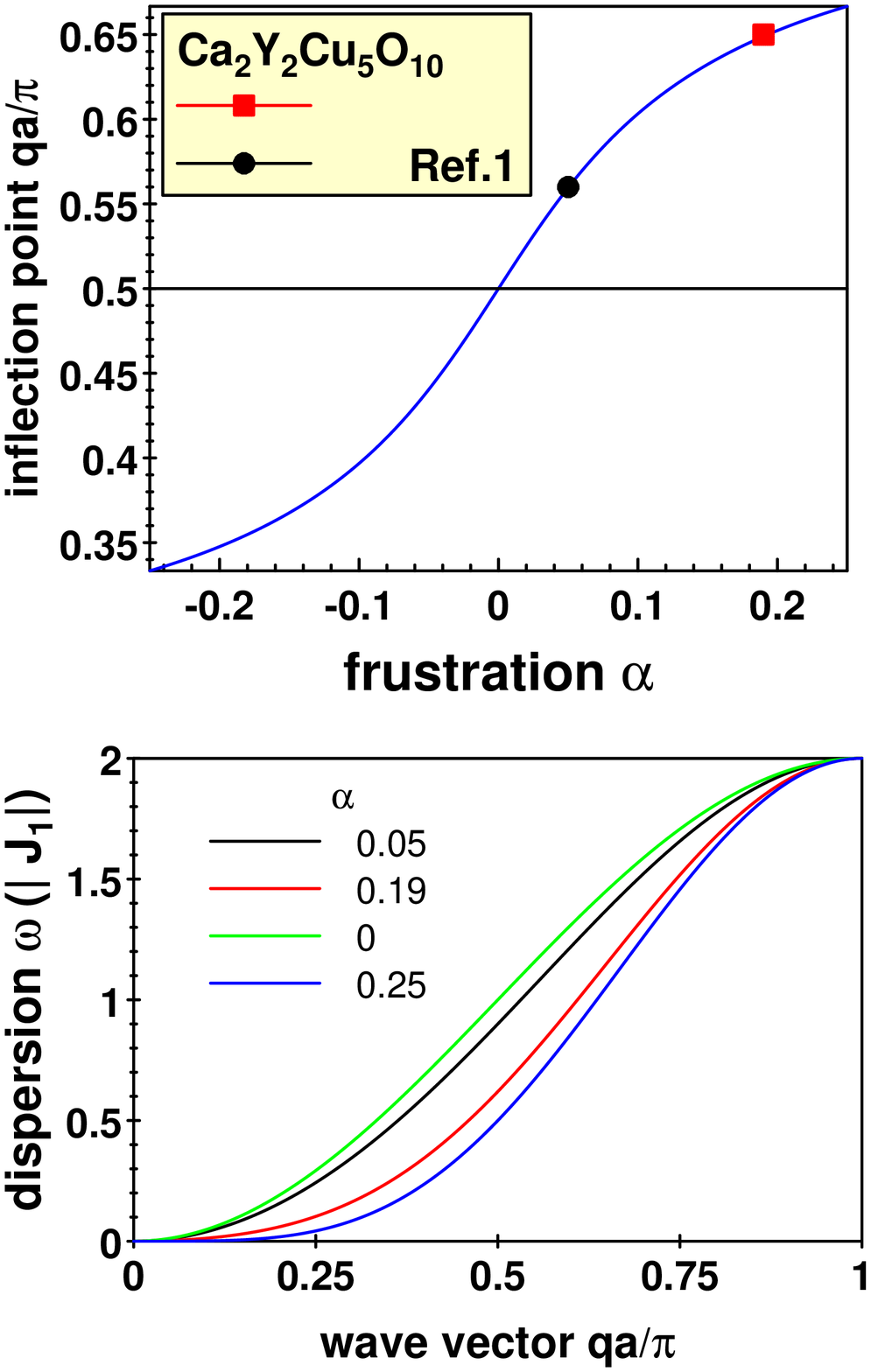}
\caption{(Color)  The influence of the frustration $\alpha$ on the dispersion
of magnetic excitation for a direction not affected by the interchain coupling.
e.g.\ (H,0,1.5) (see Fig.\ 2 of the main text).
Uper: position of the inflection point. Blue curve Eq.\ (S25).
lower: the full dispersion for various $\alpha$-values.
}
\label{frust1} 
\end{figure}
\begin{figure}[t]
\includegraphics[width=0.7\columnwidth]{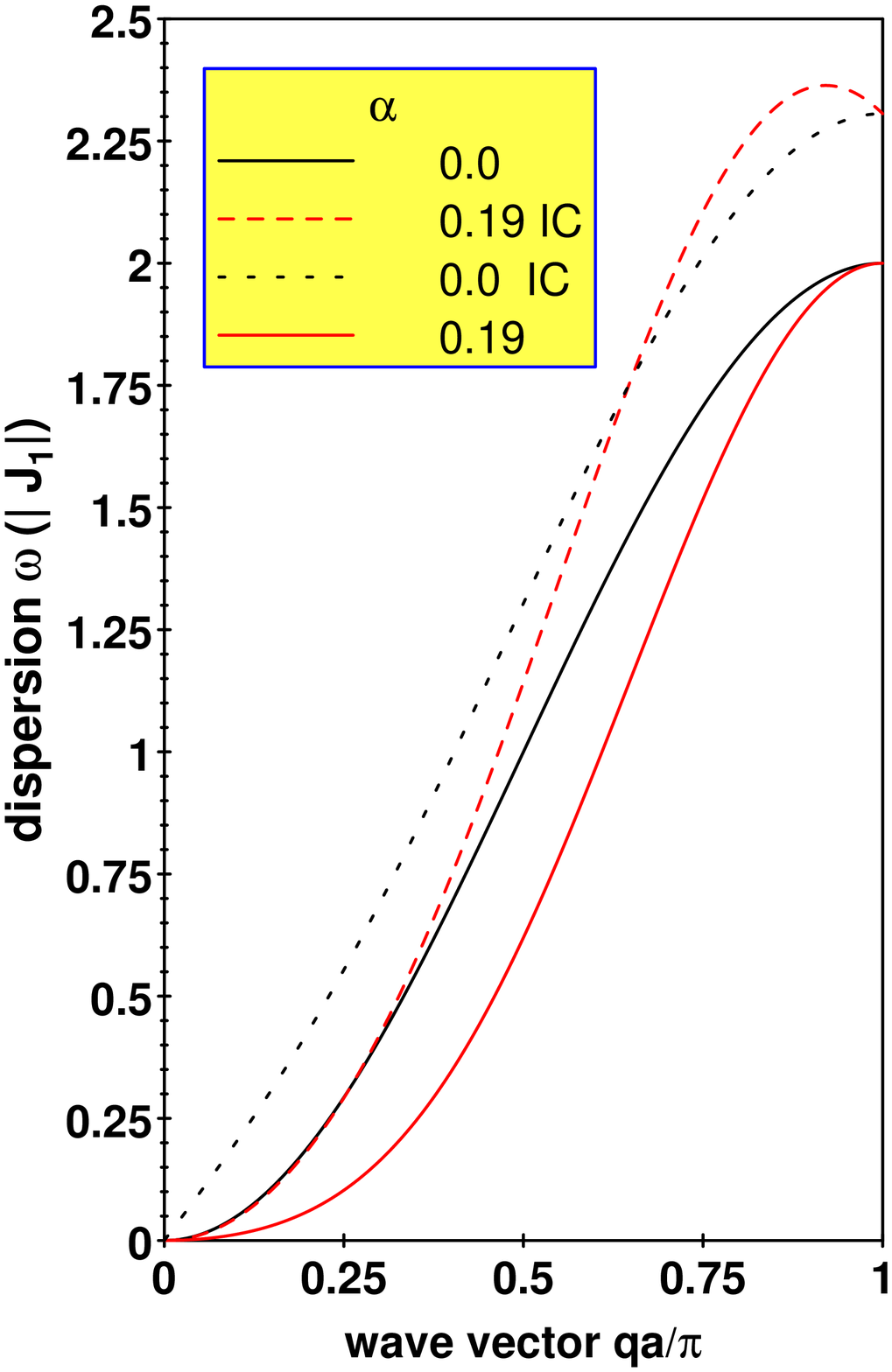}
\caption{(Color)  The influence of the frustration $\alpha$ on the dispersion
of magnetic excitation for a direction affected by the interchain coupling
(IC) 
e.g.\ (H,0,0) (see Figs.\ 2 and 3 of the main text).
Black curves: no inchain frustration, red curves: inchain frustration
included. Broken curves curves:  $J_{ic}$ included,
but we adopt $J'_{ic}$=0 for the sake of simplicity.}
\label{frust2} 
\end{figure}
of the dispersion curve in the case under consideration. 
According to our results of the mapping of the 5-band extended Hubbard
$pd$-model on a frustrated spin model $\gamma$ is usually FM
but very small $\sim 10^{-2}$ to $10^{-3}$. In many cases it can be  
therefore neglected.
The whole width of the dispersion $W$
is given by the odd Fourier coefficients, only:
$ W = 2 (\mid J_1 \mid  +J_3 + ...)$.
In the limit $x \ll 1$ Eq.\ (S22) can be rewritten as
\begin{equation}
\frac{\omega}{\mid J_1 \mid}-2\beta \approx 
\frac{1-4\alpha -9\gamma}{2}x^2
-\frac{1-16\alpha -81\gamma}{24}x^4 +O(x^6) \ .
\end{equation}
Thus, the inchain frustration  $\alpha >0$ reduces the quadratic term 
(which vanishes approaching the critical point, see Fig.\ S1) 
and strongly affects the quartic term which is
{\it negative} for the unfrustrated FM
($\alpha=0$) but changes it sign
for a sizable value of $\alpha > (1-81\gamma )/16 \approx $ 1/16 to 1/8, 
well realized for CYCO with $\alpha=0.19$. Thus, a {\it positive} sign of 
the 
quartic term gives direct evidence for a pronounced
frustrated ferromagnet.
Just the opposite behavior is realized near the maximum of the dispersion
near the BZ boundary. 
There the expansion in terms of $\varepsilon =\pi -qa$
reads
\begin{eqnarray}
\frac{\omega}{\mid J_1 \mid}-2\beta \approx 
2(1-\gamma) -
\frac{1+4\alpha -9\gamma}{2}x^2+\nonumber \\
+\frac{1+16\alpha -81\gamma}{24}x^4 
+O(x^6), 
\end{eqnarray}
the inchain frustration $\alpha$
enhances the negative
quadratic term (see also Fig.\ S1)
In the case $J_3 = 0 $ approximately fulfilled
for almost all ELC, 
the inflection point $q_{ip}$  of the dispersion curve (Eq.\ (S22)) is  
given by
\begin{equation}
q_{ip}a=\cos ^{-1}\left( \frac{1-\sqrt{1+128\alpha^2}}{16 \alpha } \right) . 
\end{equation}
It 
provides also a clear measure of the inchain frustration  
$\alpha =J_2/\mid J_1 \mid$ 
present in the system (see Fig.\ S1).
To summarize, analyzing the full dispersion curve along the line
(H,0,1.5) the presence of the inchain frustration $\alpha$
 is easily detected.

However, we admit that in many cases where the full dispersion curve is not available,
a fit of a significant part of the total curve might provide a more
precise value of $\alpha$ or $J_2$. In such cases the extraction of $\alpha$
from the $T$-dependence of the Zhang-Rice exciton intensity
shown in the main text might 
provide a reasonable alternative.

Now we briefly consider the (H,0,0) 
 scattering direction where the interchain  coupling (IC) 
is involved. Compared with the former "1D"-case we show in Fig.\  S2
that the IC has a sizable influence on the shape and on the curvature
of the dispersion. Note 
that 
a small quadratic term exists only for finite anisotropy $-D>0$.
\begin{equation}
\frac{\omega}{\mid J_1 \mid}=\sqrt{ 
\Delta^2_0 +\frac{1}{2} [ (1-4\alpha ) \beta_D 
+2\beta ( \beta +8\beta_2  ) ] x^2 +O(x^4)  }, 
\end{equation}
where $\delta=-D/|J_1|$, $\beta=J_s/\mid J_1\mid$, 
and $\beta_2=J_{ic}/|\mid J_1\mid $.
In the isotropic limit $D$ or $\delta=0$, i.e.\ when the spin gap
$\Delta_0=\sqrt{4\delta\beta +\delta^2}$ vanishes, 
 the dispersion becomes linear (compare Fig.\ 3 (left) in the main text).

Like in the former case one observes that the 
in-chain frustration $\alpha$ strongly enhances the higher order terms near the 
$\Gamma$ point. But the interchain coupling reduces this enhancement.
Under such circumstances the extraction of the IC and $\alpha$ from a 
a scattering direction affected by both types of interactions 
is unconvenient.
\subsection*{Magnetic excitations within the model of coupled chains}
We fix the sum of the two interchain couplings $J_s=J'_{ic}+J_{ic}=2.241$~meV
at the value derived from
the INS data and change only their ratio $\tau$.
First we switch off the frustrating NNN AFM
inchain coupling
$J_2$ and adopt for $J_1$ the value suggested in 
Ref.\ \onlinecite{Matsuda01} (see Fig.\ 5)
One observes roughly two kinds of dispersing peaks:
one with a dominant intensity well discribed by the LSWT
and second one which much less intensity.
The latter 
type of
curves
has been interpreted as "evidence" for an additional anomalous
second ME
branch in Refs.\ \onlinecite{Matsuda01}.
A similar behavior is observed for all cases with a dominant NN interchain coupling 
$J'_
{ic}$. In the opposite case of a dominant $J_{ic}$ that second curve almost 
invisible. For our parameter set that minority peaks are much less pronounced.
Thus we conclude, that this quantum effect depends on the details of the two main AFM interchain
couplings and becomes weaker with increasing $\mid J_1 \mid$ in contrast to
what has been suggested previously. Anyhow,
a systematic study including the examination of finite size scaling is left
for a future investigation since it is of less relevance for CYCO.
   
\begin{figure}[]
\includegraphics[width=0.7\columnwidth ]{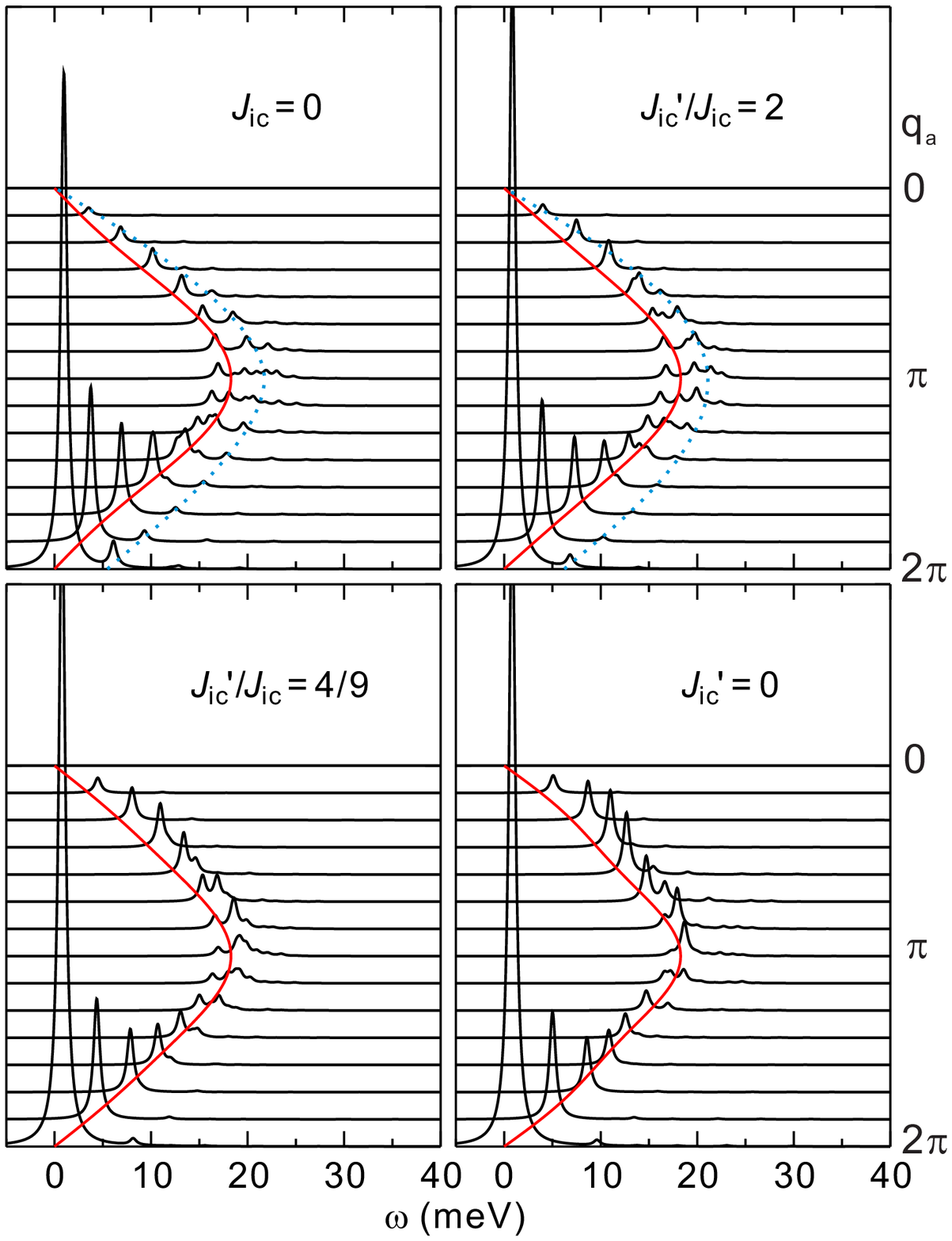}
\caption{The dynamical structure factor from exact diagonalizations
of two coupled chains with 14 sites for each chain 
at various energies $\omega$ and momenta $q$.
Red line:
projected maximum position of $S(\omega, q)$
very close to the LSWT result. 
Blue dashed line: intensity
from 
the weak 
minority peaks 
and the inchain parameters proposed in Ref.\ 1.
}
\label{fig1} 
\end{figure}
\begin{figure}[]
\includegraphics[width=0.7\columnwidth ]{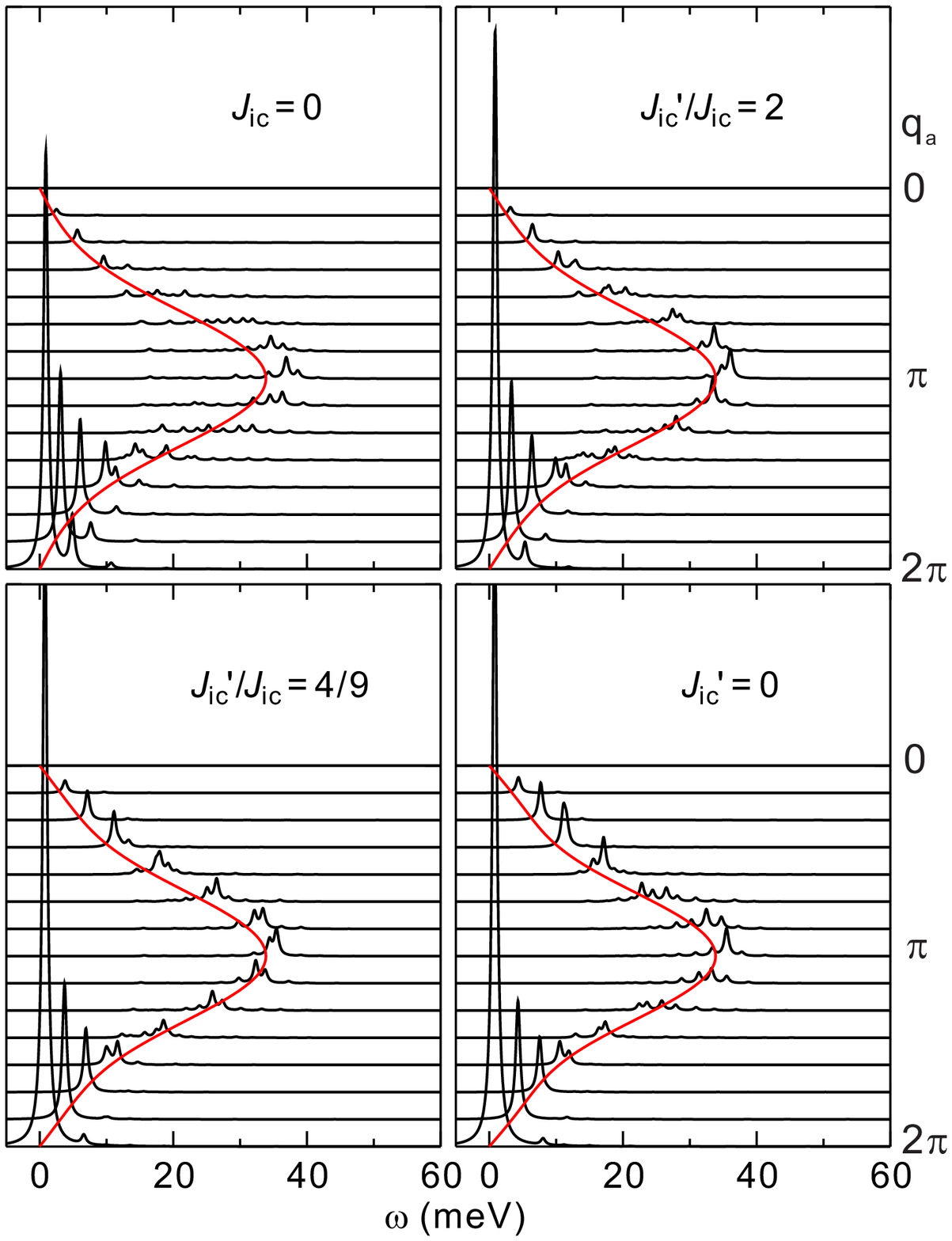}
\caption{(Color) The same as in Fig.\ S3 for the new
parameter set given in the main text.
}
\label{fig1} 
\end{figure}
\subsection*{The reduction of the magnetic moment at $T=0$}
In contrast to the case of FM ordering, the vacuum state
for the antiferro-magnons
$\alpha_{\mathbf{q}},\beta_{\mathbf{q}}$ (\ref{eq:uv}) does not
coincide with the classical Neel ground state 
$\left|\mathrm{Neel}\right\rangle \neq\left|0\right\rangle $.
Thus, the vacuum $\left|0\right\rangle $ contains 
an 
finite number
of spin deviations (\ref{nmA}),(\ref{nmB}) even at 
$T=0$, 
$\left\langle 0\right|\hat{n}_{\mathbf{m}}\left|0\right\rangle \neq0$. 
Expressing the 
spin deviation operators $a_{\mathbf{q}}$ via 
$\alpha_{\mathbf{q}},\beta_{\mathbf{q}}$
(\ref{eq:uv}), we obtain 
\begin{eqnarray}
& & \left\langle 0\right|a_{\mathbf{m}}^{\dagger}a_{\mathbf{m}}
\left|0\right\rangle   =  
\frac{2}{N}\sum_{\mathbf{q}}
\left\langle 0\right|a_{\mathbf{q}}^{\dagger}a_{\mathbf{q}}
\left|0\right\rangle \nonumber \\
&=& \frac{2}{N}\sum_{\mathbf{q}}\sinh^{2}\theta_{\mathbf{q}}
\left\langle 0\right|\beta_{\mathbf{q}}\beta_{\mathbf{q}}^{\dagger}
\left|0\right\rangle   =  
\frac{1}{N}\sum_{\mathbf{q}}
\left(\frac{A_{\mathbf{q}}}{\omega_{\mathbf{q}}}-1\right). \nonumber \\
\end{eqnarray}
Note, that in our geometry, the summation over $\mathbf{q}$ runs
over the magnetic Brillouin zone (BZ) which
in the present case  coincides with the lattice 
BZ $\frac{2\pi}{a}\times\frac{2\pi}{b}\times\frac{2\pi}{c}$.
Finally, for the sublattice magnetization at $T=0$, we have
\begin{equation}
\left\langle 0\right|\hat{S}_{\mathbf{m}}^{z}\left|0\right\rangle =
S\left(1-\frac{\left\langle 0\right|\hat{n}_{\mathbf{m}}
\left|0\right\rangle }{S}\right). \label{m}
\end{equation}
\subsection*{ The optical conductivity within the 5-band extended Hubbard model}
Focusing on the Zhang-Rice exciton,  we note that a similar picture as
in the RIXS spectra is shown in the predicted optical conductivity
and EELS spectra (not shown here). Some insight in the corresponding transitions 
can be gained from
the zoomed figure with an artificial small broadening 
to resolve these transitions
(see Fig.\ S5).
\begin{figure}[b!]
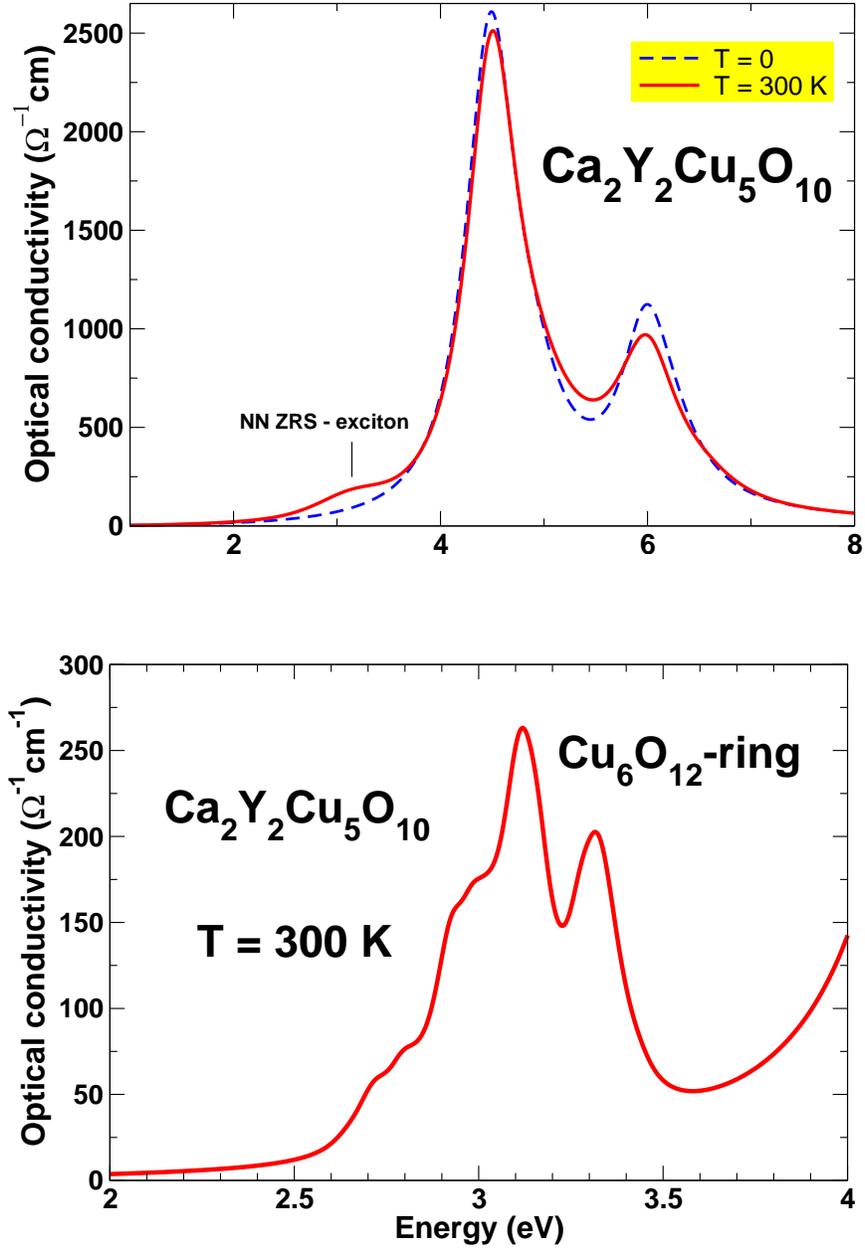

\includegraphics[width=0.65
\columnwidth]{sigma.eps}
\includegraphics[width=0.65
\columnwidth]{opthexaZRS300K.eps}
\caption{(Color) The optical conductivity $\sigma (\omega)$
for a Cu$_6$O$_{12}$ cluster
with periodic boundary conditions
within the  5-band extended Hubbard Cu 3$d$ O$2p$
model and parameters given
in the main text. Upper: Broad 
energy
region and
spectrum
broadened by 
$\Gamma =0.3$~eV. Notice the NN Zhang-Rice exciton visible in 
the
spectrum at 300~K
and its lacking at $T=0$ (compare Fig.\ 5. in the main text).
Lower: The same as in the upper  for
the region of Zhang-Rice singlet transitions.
The spectrum has been broadened by 
$\Gamma =0.05$~eV, only, to make all transitions visible. 
Notice the main Zhang-Rice singlet exciton near
3~eV.
}
\label{figopt} 
\end{figure}
\begin{table}
\begin{ruledtabular}
\begin{tabular}{l  c c c}
site & x/a & y/b & z/c\\
 \hline
Cu& 3/8&0&1/4\\
Cu&7/8&0&1/4\\
Cu&1/8&1/2&1/4\\
Cu&5/8&1/2&1/4\\
O&3/8&0&.6270\\
O&7/8&0&.6270\\
O&1/8&1/2&.6270\\
O&5/8&1/2&.6270\\
O&3/8&0&.8730\\
  O&7/8&0&.8730\\
 O&1/8&1/2&.8730\\
O&5/8&1/2&.8730\\
Na&1/8&1/4&0\\
   Na & 5/8&1/4&0\\
      Na & 1/8 &1/4 &1/2\\
      Na&5/8 &1/4 &1/2\\
\end{tabular}
\end{ruledtabular}
\caption{\label{struc} Crystal structure of the {\it 
commensurate }
approximate 
effective
Na$_2$CuO$_2$ compound 
for DFT+$U$ calculation (enlarged unit cell and reduced symmetry to allow different spin configuration).
Space group P2/M (SG 10), $a=5.6306~$\AA, $b=6.286~$\AA\ and $c=10.5775~$\AA. }
\end{table}
\subsection*{ Additional information on the L(S)DA+$U$ calculations}
To model the mutually incommensurate crystal structure of CYCO we neglect 
(i) the modulation of the Cu-O distances within the CuO$_2$ chains and 
(ii) the incommensurability of the CuO$_2$ and CaY subsystems as a good 
approximate to estimate the NN exchange along the CuO$_2$ chains. 
The crystal structure data for this simplified model structure are given in 
Tab.~S2. To calculate the NN exchange $J_1$ for the approximated crystal 
structure we constructed a super cell by the doubling of the unit cell and
the reduction of its symmetry to allow for different spin configurations 
(see Tab. ~S1). 
The obtained FM NN exchange $J_1$ depending on the Coulomb repulsion 
$U_{3d}$ and the specific functional are depicted in Fig.~\ref{S5}. $J_1$ 
depends only weakly on these parameters. The given NN exchange
$J_1$=-150\,K is the average between the LDA and GGA result at $U_{3d}$=6.5\,eV.
\begin{figure}[]
\begin{center}\includegraphics[%
  clip,
  width=5.5cm,
  angle=0]{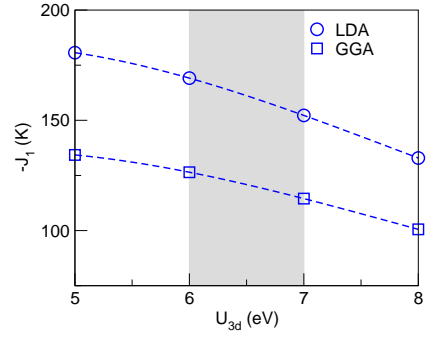}\end{center}
\caption{\label{S5}(Color online)
Calculated FM NN exchange $J_1$ as function of the Coulomb repulsion $U_{3d}$.
}
\end{figure}
\begin{table}[b!]
\begin{ruledtabular}
\begin{tabular}{l  c c c}
site & x/a & y/b & z/c\\
 \hline
Cu &0 & 0& 0\\
O & 0& 0& 0.623\\
Na & 1/2& 1/4&1/4 \\
\end{tabular}
\end{ruledtabular}
\caption{\label{struc} Crystal structure of the {\it commensurate} approximate 
effective Na$_2$CuO$_2$ structure as a starting model for CYCO.
Space group FMMM (SG 69), $a=2.8153$~\AA, $b=6.286$~\AA\ and $c=10.5775$~\AA. 
}
\label{struc2}
\end{table}
\subsection*{ Aspects of the magnetic susceptibility}
In addition to 1D susceptibilities,
for the isotropic as well as the easy-axis anisotropic case
 calculated using the transfer matrix renormaliztion group theory
method, also
the  3D case has been examined
treating the adopted isotropic
interchain coupling (IC)
within the RPA (random phase approximation) (see Fig.\ S7 and Eq.\ (S29)).
\begin{equation}
\chi_{\mbox{\tiny 3D}}(T)\approx 
\frac{
\chi_{\mbox{\tiny 1D}}(T)}{1+k\chi_{\mbox{\tiny 1D}}(T)} \quad .
\end{equation}
\begin{figure}[t!]
\includegraphics[clip,width=0.47\columnwidth]{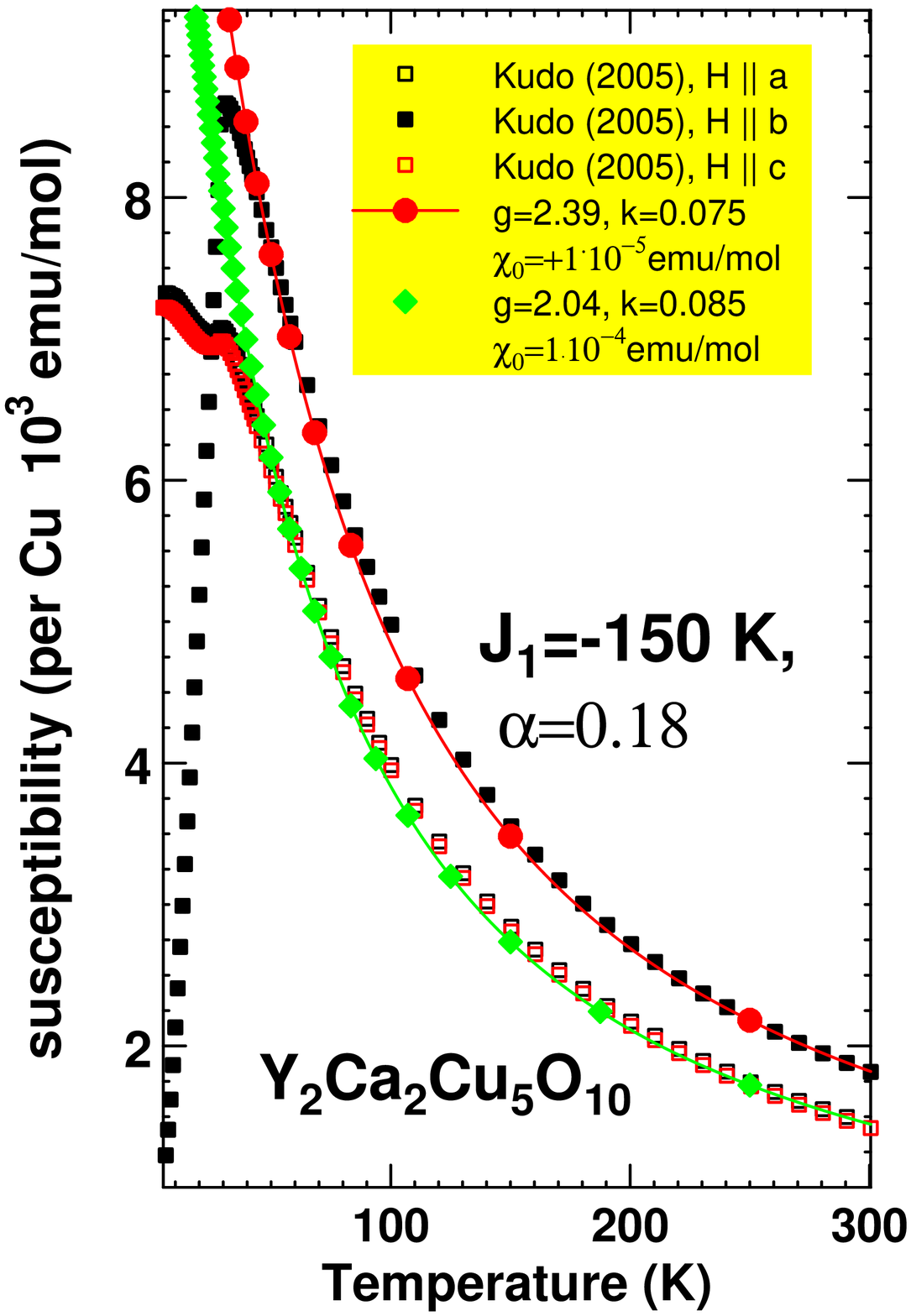}
\includegraphics[clip,width=0.47\columnwidth]{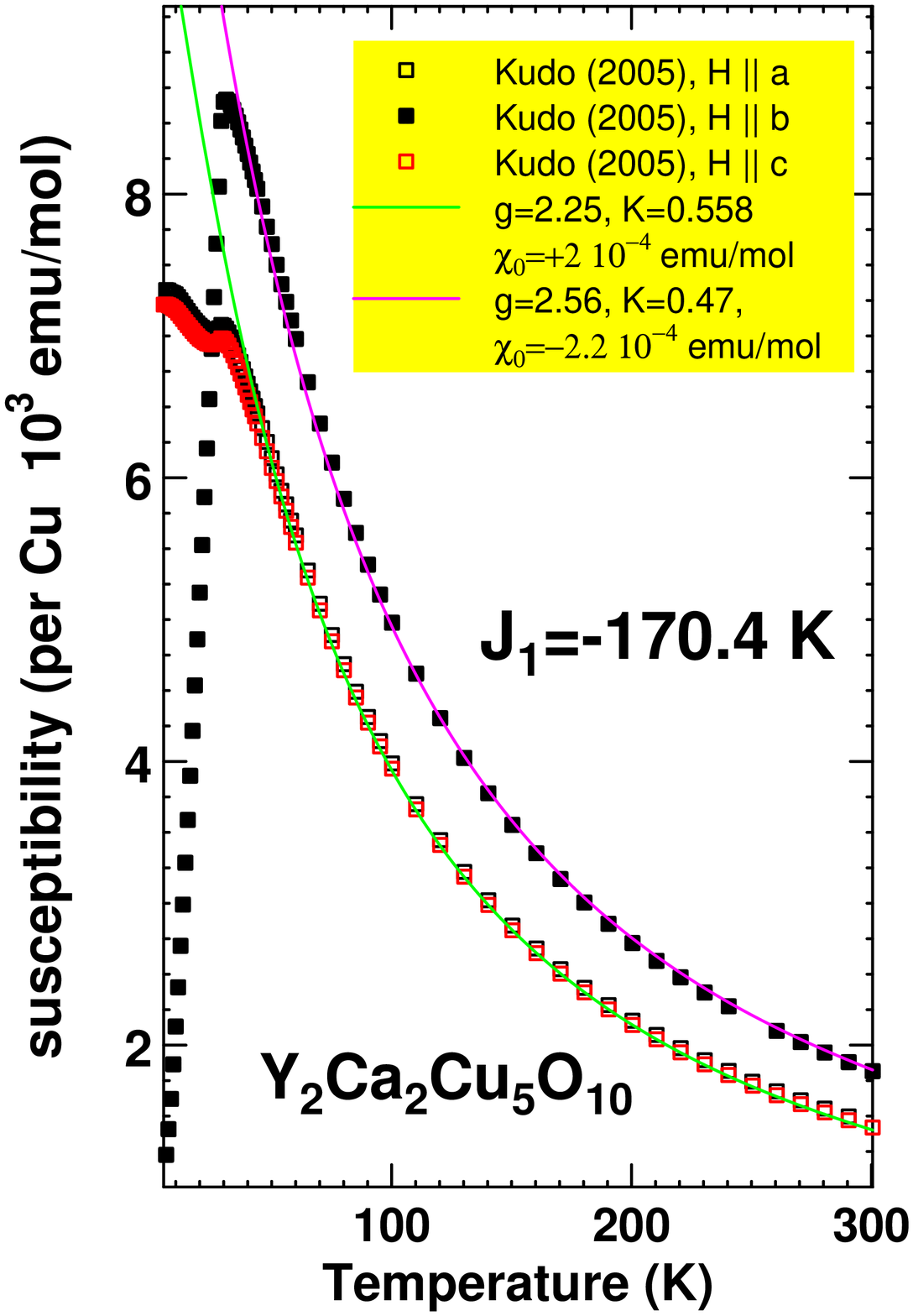}
\caption{
(Color) Left: spin
susceptibility $\chi(T)$ fitted within
the 1D $J_1$-$J_2$ model  supplemented with isotropic AFM IC in 2D
 treated within the RPA . The latter is measured by the parameter 
$k=2\left( J'_{ic }+J_{ic} \right)/\mid J_1\mid $.
Right: The same as left for an
anisotropic easy-axis
inchain coupling $J_1$.
}
\label{figchi1} 
\end{figure}

\vspace{0.3cm}
\noindent
$[1]$ M.\ Matsuda   {\it et al.}, Phys.\ Rev.\ B, {\bf 63},
180403 (2001). \\
$[2]$ T.\ Oguchi, Phys.\ Rev.\ {\bf 117}, 117 (1960).\\
\end{widetext}

\begin{thebibliography}{99}
\bibitem{remarkparameters}
We used $U_d=8.5$~eV, $U_p=4.1$~eV,
the interorbital onsite Coulomb repulsion on
O sites $U_{p_xp_y}=2.9$~eV, charge transfer energies 
$\Delta_{p_xd}=3.5$~eV, 
and $\Delta_{p_yd}=3.8$~eV (close to Ref.\ \onlinecite{Mizuno98} 
with an isotropic value $\Delta_{pd}=3.8$~eV).  We further include 
the intersite Coulomb interaction $V_{pd}=0.5$~eV
neglected in Ref. \onlinecite{Mizuno98}. This  
seems to be the main reason for the difference
of the ZRS-exciton energies, upshifted by $\sim$ 0.3~eV
compared to Ref.\ \onlinecite{Mizuno98}.
\end{thebibliography}
\end{document}